\documentclass[aip,cha,numerical,showkeys,showpacs,groupedaddress,amsmath,amssymb,floatfix,twocolumn,reprint]{revtex4-1}

\usepackage[utf8]{inputenc}
\usepackage{subcaption}
\usepackage{amsmath}
\usepackage{amssymb}
\usepackage{lmodern}
\usepackage[pdftex]{graphicx}
\usepackage{hyperref}

\usepackage{changes}
\definechangesauthor[name={Frank}, color=red]{FH}
\definechangesauthor[name={Paul}, color=green]{PS}
\definechangesauthor[name={Kevin}, color=blue]{KW}

\newcommand{\dd}{\mathrm{d}}
\newcommand{\ind}[1]{\mathbf{1}_{#1}}

\newcommand{\br}[1]{\mathopen{}\left( #1 \right)\mathclose{}}
\renewcommand{\exp}[1]{\mathrm{e}^{#1}}

\newcommand{\In}{\iota}
\newcommand{\BS}{\beta}
\newcommand{\FTBS}{\BS}

\begin{document}

\title{Bounding the first exit from the basin:\\Independence Times and Finite-Time Basin Stability}

\author{Paul~Schultz$^{1,2}$}
\author{Frank~Hellmann$^{1}$}
\author{Kevin~N.~Webster$^{1}$}
\author{J\"{u}rgen~Kurths$^{1,2,4,5}$}


\email{pschultz@pik-potsdam.de}

\affiliation{$^1$Potsdam Institute for Climate Impact Research (PIK), Member of the Leibniz Association, P.O. Box 60 12 03, D-14412 Potsdam, Germany\\
$^2$Department of Physics, Humboldt University of Berlin, Newtonstr. 15, 12489 Berlin, Germany\\
$^3$Institute for Complex Systems and Mathematical Biology, University of Aberdeen, Aberdeen AB24 3UE, United Kingdom\\
$^4$Department of Control Theory, Nizhny Novgorod State University, Gagarin Avenue 23, 606950 Nizhny Novgorod, Russia}

\begin{abstract}
We study the stability of deterministic systems given sequences of large, jump-like perturbations.
Our main result is to dervie a lower bound for the probability of the system to remain in the basin, given that perturbations are rare enough.
This bound is efficient to evaluate numerically.

To quantify rare enough, we define the notion of the independence time of such a system. This is the time after which 
a perturbed state has probably returned close to the attractor,
meaning that subsequent perturbations can be considered separately.
The effect of jump-like perturbations that occur at least the independence time apart is thus well described by a fixed probability to exit the basin at each jump, allowing us to obtain the bound.

To determine the independence time, we introduce the concept of finite-time basin stability,
which corresponds to the probability that a perturbed trajectory returns to an attractor within a given time. 
The independence time can then be determined as the time scale at which the finite-time basin stability 
reaches its asymptotic value. Besides that, finite-time basin stability is a novel probabilistic stability
measure on its own, with potential broad applications in complex systems.
%
\end{abstract}

\maketitle

\section{Introduction}\label{sec:intro}

A typical problem in the study of multi-stable dynamical systems is 
the stability of an attractor against perturbations. 
For small perturbations, stability can be assessed in terms of asymptotic
stability theory for linear systems \cite{Lyapunov1907}, 
e.g. by calculating Lyapunov exponents.

On the other hand, for large perturbations, a typical approach is
to assess stability by properties of the basin of attraction, for instance 
their size \cite{Wiley2006,Klinshov2015,Mitra2015}. For this, several direct and sampling-based
methods are available.  

In particular, Lyapunov functions \cite{Hahn1958,Malisoff2009,Giesl2015} 
and related concepts like non-equilibrium potentials \cite{Graham1984,Graham1991}
are powerful tools for studying basins of attraction.
The existence of a global Lyapunov function ensures global stability against all perturbations. 

The explicit construction of Lyapunov functions for a given system is a difficult problem in general. 
However, several numerical approaches for the computation of Lyapunov functions have been developed, 
including the SOS (sums of squares) method \cite{Parrilo2000}, the CPA (continuous piece-wise affine) 
method \cite{Hafstein2004}, radial basis functions \cite{Giesl2007}, and the numerical solution of Zubov's equation \cite{Camilli2001}. 
For a survey of these methods, see \cite{Giesl2015}.

Direct methods, however, are typically not efficient for high-dimensional systems
and yield conservative bounds on the attraction basin \cite{Chiang2010,Gajduk2014}.
 
Basin stability $\BS$ \cite{Menck2013,Schultz2017,Mitra2017} instead studies the probability 
that a system will return to an attractor following a large, jump-like perturbation. 
As other measures designed this way (e.g. \cite{Rega2005,Hellmann2016}), 
it has the advantage of allowing for efficient estimators by sampling the phase space and the trajectories directly. 
These estimators have a sampling error that is independent of the system dimension. 
Thus $\BS$ can be efficiently evaluated for high-dimensional systems and for dynamics where no 
analytic Lyapunov functions are known \cite{Schultz2017}.

In this paper, we study the behaviour of systems under repeated large perturbations.
We answer the question of how rare perturbations need to be for basin stability to predict its probability to remain in the basin indefinitely.

To do so we introduce the notion of the independence time of a system subject to a random perturbation. This captures the time the system takes to return to the attractor following a perturbation.
An inescapable problem when studying the return of a system to an attractor lies in the fact that this return typically 
takes infinitely long and requires regularisation \cite{Kittel2016,Mitra2017a}. 
Here we make use of the repeated perturbations to provide us with a meaningful regulator.
We consider the system to have returned if the dynamics have erased the memory of the previous perturbation.
More formally, the system has returned, if its distribution following a perturbation is
approximately equal to its distribution after a perturbation centered on the attractor.
If this is the case, the states after subsequent perturbations, considered as random variables, are approximately independent, and the probability to exit the basin factorizes.

To efficiently evaluate the independence time, we introduce the notion of finite-time basin stability $\FTBS\br{T}$. 
This is a finite-time horizon version of basin stability, corresponding to the probability that a system has returned to the attractor 
(according to a chosen criterion) in time $T$. By combining this with the return criterion required for independence time we can give 
a lower bound for the independence time as the time when the finite-time basin stability approaches its asymptotic value.
Furthermore, this enables us to derive an efficient estimator for a lower bound on the independence time for high-dimensional systems.

Given a set of perturbations that occur less frequently than the independence time, the probability to exit the basin of attraction 
is simply given in terms of the basin stability and the frequency of perturbations.
This is particularly of interest if the asymptotic basin stability is close to unity for a given set of perturbations.
Then, the independence time is the time interval that has to pass between perturbations to ensure that a 
sequence of such perturbations can not destabilise the system.

\section{Definitions} \label{sec:setup}

\subsection{The system}
We will consider an autonomous multi-stable dynamical system for which we can describe the dynamical 
evolution with a system of first-order ordinary differential equations, i.e.

\begin{align} \label{eq:system_no_pert}
\begin{split}
\dot{x} = f\br{x}\;,
\end{split}
\end{align}

with states $x$ living in a phase space $X\subseteq\mathbb{R}^n$. 
We are interested in the case that the system has at least one stable fixed point, 
which, without loss of generality, we assume to be at the origin $x^\ast = 0$, such that $f(x^\ast) = 0$. 
We denote the basin of attraction of the origin as $B \subset X$. Accordingly, the \emph{basin stability}~\cite{Menck2013}
of the fixed point $x^\ast$ with respect to a probability density $\rho$ of perturbations is given by 

\begin{align} \label{eq:bs}
\begin{split}
\BS := \int\limits_X \ind{B}\br{x} \rho\br{x} ~d x ,\quad \BS \in [0;1] \;.
\end{split}
\end{align}

$\BS$ corresponds to the probability that the system -- initially at $x^\ast$ -- returns to the 
fixed point for a perturbation drawn from $\rho$. It is proportional to the basin volume if $\rho$ 
is chosen as a uniform probability density with large enough support.

We now subject the system of Eq.~\ref{eq:system_no_pert} to a possibly infinite sequence of jump 
perturbations, with magnitude $\Delta x_i$ drawn at random from a probability density $\rho(\Delta x)$ and starting 
at time $t = 0$. We do not further specify the discrete times $t_i$ at which these perturbations 
occur, i.e. perturbations might appear regularly or according to some distribution. 
The minimum difference between subsequent perturbations will be denoted by $\Delta t = \min_i\br{t_i - t_{i + 1}}$. 
Initialising the system at the attractor, this setup leads to the stochastic integral equation

\begin{align} \label{eq:system}
\begin{split}
x(t) = \int_0^t\dd t' f\br{x\br{t'}}  + \int_0^t\dd t' \sum_{i=0}^\infty  \Delta x_i \delta\br{t' - t_i} \;.
\end{split}
\end{align}

For convenience we introduce the \emph{number of jumps} $n(t)$ that have happened at a time $t$:

\begin{align} \label{eq:number_of_jumps}
\begin{split}
n(t) = \int_0^t\dd t \sum_{i=0}^\infty \delta\br{t - t_i}\;.
\end{split}
\end{align}

We will be concerned with the remain probability
\begin{align}
P_{remain}(t, x(0)) := P\br{\forall\; 0 \leq t' \leq t:\; x\br{t'} \in B}\;.
\end{align}
That is, the probability for the system to continuously remain within the basin of attraction. This is the cumulative probability of the complement of the distribution of the time of the first exit from the basin $p_{fe}(t)$ of the system:

\begin{align}
P_{remain}(t, x(0)) = 1 - \int_0^t p_{fe}(t') dt'\;.
\end{align}

Hence all information about the exit times, including escape rates, is captured by it.

If the jumps in the system are sufficiently rare we expect that the probability for a solution $x\br{t'}$ 
to Eq.~\eqref{eq:system} to continuously remain in the basin of attraction up to time $t$ to be given by

\begin{align} \label{eq:remain-probability}
\begin{split}
P_{remain}(t, x(0)) \approx \BS^{n(t)}\;,
\end{split}
\end{align}

that is, every perturbation counted by $n(t)$ has an equal and constant probability to leave the system within
the basin of attraction (or for pushing it out). 

In Sec.~\ref{sec:independence} we will quantify what sufficiently 
rare means to achieve such a formula. Before, as an additional prerequisite, we turn to the definition of 
finite-time basin stability.

\subsection{Finite-Time Basin Stability} 

Our analysis is based on the return times of perturbed states within the basin of attraction $B$
to the original attractor, i.e. to the fixed point $x^\ast$, defined through a time-tracking Lyapunov function.
A \emph{Lyapunov function} is a function $V(x)$ with negative orbital derivative, i.e. 
it decreases along trajectories of Eq. \ref{eq:system_no_pert} and has a minimum or diverges to $- \infty$ at the 
fixed point $x^\ast$ \cite{Lyapunov1907,Hahn1958,Malisoff2009}. Further, the fixed point is the only point in the basin for which it is minimal or negatively divergent.
Given that such a function  exists and $f(x)$ is sufficiently smooth, $x^\ast$ is asymptotically stable.
A \emph{time-tracking Lyapunov function} $V$ is defined on $B$ and satisfies the differential equation

\begin{align} 
\begin{split}
\frac{d}{dt} V\br{x(t)} = -1\;,
\end{split}
\end{align}

i.e., it strictly decreases along any trajectory of Eq.~\eqref{eq:system}.
It is straightforward to see that the values of such Lyapunov functions track the time. 
If $x(t)$ and $x(t^\prime)$ are two points on the same trajectory, then by integrating the defining 
equation above we have:

\begin{align} 
\begin{split}
V(x(t)) - V(x(t^\prime)) = t - t^\prime\;.
\end{split}
\end{align}

To fully determine such a Lyapunov function we need to specify boundary conditions on a transverse surface $S$ (more precisely we require the surface to be non-characteristic, see, e.g. \cite{Giesl2007}).
If we set $V(S) = 0$, the time-tracking Lyapunov function measures how long it has been since, or will be until the system crosses the surface $S$. 
We denote this Lyapunov function as $V_S$. We further assume that $S$ lies in $B$ entirely, and set $V_S\br{x}=\infty$ for states $x\in X\setminus B$ outside the basin of attraction.

The set $S$ defines our return condition and the \emph{finite-time basin stability}, 
given $\rho$ and $S$, is defined as:

\begin{align} \label{eq:ftbs}
\begin{split}
\FTBS_S(T) := \int\limits_X \ind{B}\br{x} \Theta\left(T - V_S(x)\right) \rho\br{x} ~d x \;\in [0;1]\;.
\end{split}
\end{align}

$\Theta$ denotes the Heaviside step-function. This is the probability that a trajectory, following a perturbation drawn from $\rho(x)$,
will return to within $S$ around the attractor $x^\ast=0$ within time $T$. For well-behaved vector fields $f(x)$, one expects that 
$\lim_{T \rightarrow \infty} \FTBS_S(T) = \BS$.  Note that the latter does not assume $S$ to be small but holds for all $S\in B$ even
if they enclose almost the whole basin.

\section{Approximate Independence of post perturbation states}\label{sec:independence}

\begin{figure}[ht!]
    \centering
    \includegraphics[width=.8\columnwidth]{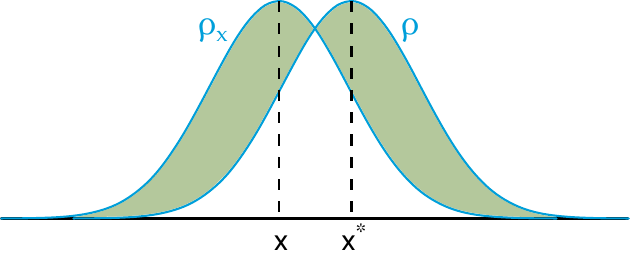}
    \caption{ 
    \textbf{Schematic representation of $\In\br{x, x^\ast}$:}
    Imagine a one-dimensional system Eq.~\ref{eq:system_no_pert} with a fixed point $x^\ast$.
    The difference between the probability density $\rho$ centred at $x^\ast$ and the shifted 
    density $\rho_x$ is then given by $\In\br{x, x^\ast}$ (shaded green area) as defined in Eq.~\ref{eq:in}.
    }
   \label{fig:independence_measure}
\end{figure}

We now turn to the key question: \emph{When do we consider the system to have returned?}
As noted above, we want ``returned'' to imply that, from the current position, the state following a perturbation of the system, is 
statistically independent from the state after the preceding perturbation.

Therefore we will consider the shifted perturbation distributions.
Let us define a distance function $\In\br{x,x^\prime}$ 
on the phase space as the $L^1$ norm of the difference of the \emph{shifted probability distributions} 
$\rho_x\br{\cdot} = \rho\br{\cdot - x}$:

\begin{align} \label{eq:in}
\begin{split}
\In\br{x, x^\prime} = \int\limits_X \left|\rho_x\br{u} - \rho_{x^\prime}\br{u}\right| du\;.
\end{split}
\end{align}

This is visualised in Fig.~\ref{fig:independence_measure}, where the distance between $x$ and
the fixed point $x^\ast$ is given by $\In\br{x, x^\ast}$ as indicated by the shaded area.
Note that for an arbitrary state vector $y$, $\In\br{y, x}$ is a subadditive, symmetric, non-negative 
function of $x$ and vanishes for $x=y$, hence it is a pseudometric on $X$. 
We will use the shorthand $\In\br{x} = \In\br{x, x^\ast}$ for the distance to the fixed point.

The expectation value of some observable $\chi\br{x}$ satisfying $\left|\chi\br{x}\right| \leq 1$ 
with respect to the two distributions $\rho$ and $\rho_x$  differs at most by $\In\br{x}$:

\begin{align} \label{eq:main_approx}
\begin{split}
&\left|\int\limits_X \chi\br{u} \rho\br{u}~du - \int\limits_X \chi\br{u} \rho_x\br{u}~du\right| \\
&\leq  \int\limits_X \vert\chi\br{u}\vert \vert\rho\br{u} - \rho_x\br{u}\vert ~du \\
&\leq \int\limits_X \vert\rho\br{u} - \rho_x\br{u}\vert~du = \In\br{x}
\end{split}
\end{align}

The probability to remain in the fixed point's basin of attraction after a perturbation originating at $x$ 
is given by the basin stability $\BS_x$ of the shifted probability density $\rho_x$:

\begin{align} \label{eq:bs_shifted}
\begin{split}
\BS_x := \int\limits_X \ind{B}\br{u} \rho_x\br{u} ~d u \;\in [0;1]
\end{split}
\end{align}

Both basin stability and finite-time basin stability are defined as the expectation value of the 
basin indicator function $\ind{B}$. Thus, in particular, we have that

\begin{align} \label{eq:iota estimate}
\begin{split}
\vert\BS - \BS_x\vert \leq \In\br{x} \quad \text{and} \quad \vert\FTBS\br{T} - \BS_x(T)\vert \leq \In\br{x}\; .
\end{split}
\end{align}

For a system Eq.~\eqref{eq:system} at a jump event $t_i$, the distribution of the state after the perturbation, which we denote $x(t_i^+)$, given the state before the jump $x(t_i^-)$ is given by $\rho_{x(t_i^-)}$. Thus the difference in the probability to exit the basin from $x(t_i^-)$ as opposed to $x^*$ is bounded by $\In\br{x(t_i^-)}$. The distance to the attractor in our metric $ \In$ is a meaningful measure for the return to the attractor. If it is small, the distribution after two different jump events, $t_i$ and $t_j$, is similar, and the jumps are approximately independent in the sense we require.

%

\section{Independence Times}

\begin{figure}[!th]
    \centering
    \begin{subfigure}[t]{0.45\textwidth}
        \centering
        \includegraphics[width=0.8\columnwidth]{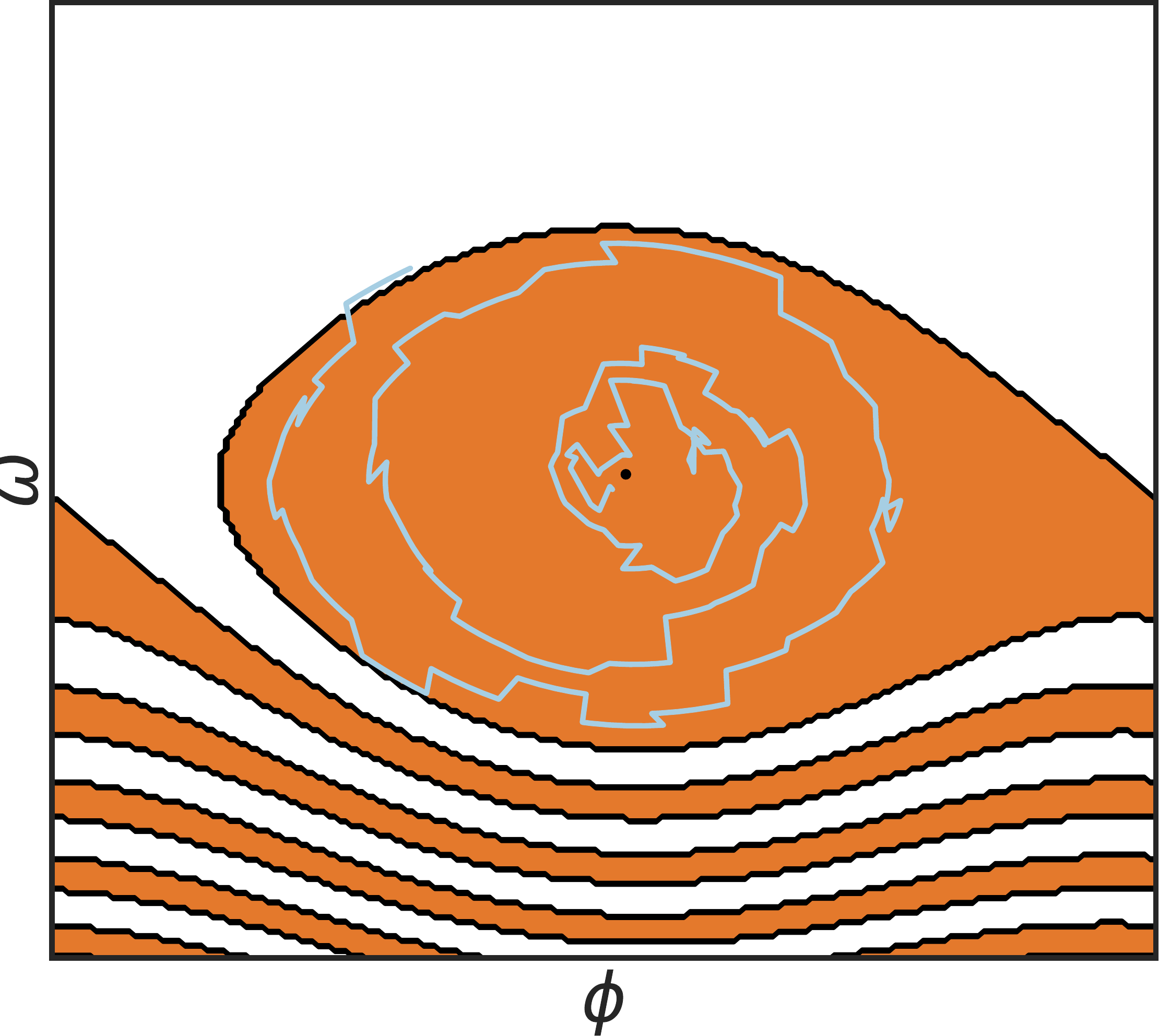}
        \caption{}
    \end{subfigure}
    
    \begin{subfigure}[t]{0.45\textwidth}
        \centering
        \includegraphics[width=.8\columnwidth]{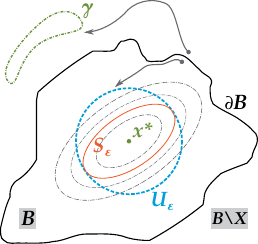}
        \caption{}
    \end{subfigure}%
   
    \caption{
    \textbf{(a)} 
    Example realisation of a swing equation  
    (Eq.~\ref{eq:swing}) describing the evolution of phase $\phi$ and frequency $\omega$
    dynamics of damped driven pendulum, discussed further in Sec.~\ref{sec:results}, subject to frequent, bounded jumps.
    The jump intervals are chosen to be comparatively short ($T=0.1$), hence 
    the trajectory quickly escapes the corresponding basin of attraction
    (orange area).
    \textbf{(b)} 
    Schematic picture of the basin of attraction $B$ with boundary $\partial B$ 
    of a the fixed point $x^\ast$ in a phase space $X$, visualising the relation of the
    sets $U_\epsilon$ and $S_\epsilon$ defined in Sec.~\ref{sec:estimator}.
    In a multistable system, trajectories either approach the fixed point or other 
    attractors, for instance a limit cycle $\gamma$.
    }
   \label{fig:schematic}
\end{figure}

To illustrate how independence can fail, consider Fig.~\ref{fig:schematic}a.
The Figure shows the phase space of a damped driven pendulum, described by phase and frequency $x=\br{\phi,\omega}$. The shaded region is the basin of attraction of the fixed point $x^\ast$ at the origin. The shown trajectory is an example realisation of the deterministic 
dynamics being subject to jump perturbations (Eq.~\ref{eq:system}) with $\Delta t$
chosen to be comparatively short. The perturbations are bounded in size, and the basin stability of the system is one. However, as they occur frequently, the system has no time to return to the attractor, leading to an eventual escape from the basin. After several jumps $\BS_x$ starts being considerably smaller than $\BS$.

We can now combine the concepts introduced above to define a time that has to pass between subsequent perturbations, in order to prevent sucha build up.

For our definition of finite-time basin stability (cf. Eq.~\ref{eq:ftbs}) we have to specify 
a transverse surface $S$ for the time-tracking Lyapunov function $V_S$. In particular, given an $\epsilon > 0$, 
we choose $S$ such that $\In\br{x} < \epsilon$ for all $x$ enclosed by $S$.
Perturbations starting from the interior of $S$ are \emph{almost} identical, with a deviation bounded by $\epsilon$.

The fact that $S$ is transverse, and it's interior points satisfy $\In\br{x} < \epsilon$, means that after the system enters $S$, $\In\br{x(t)}$ will never be larger than $\epsilon$ in the future.

Now given a threshold $\delta > 0$, we define the \emph{independence time} of a dynamical system as the time 
$T_{ind}(\epsilon, \delta)$ such that:

\begin{align} \label{eq:independence_time}
\begin{split}
T_{ind}\br{\epsilon, \delta} := \inf \{ T > 0 \vert\, \BS - \FTBS_S\br{T} \leq \delta \} \;.
\end{split}
\end{align}

That this time-scale accurately quantifies independence of subsequent perturbations for system Eq.~\eqref{eq:system} is shown by the following result:

\subsection*{Main result}
Given a sequence of perturbations drawn from $\rho$, occurring at times $t_i$ with minimum interval $\Delta t$ 
larger than the independence time $\Delta t > T_{ind}(\epsilon, \delta)$, the probability to remain within 
the basin of attraction, given that $x(0)\in S$,  is bounded by

\begin{align} \label{eq:remain-probability-epsilons}
\begin{split}
P_{remain}\br{t, x(0)} \geq \br{\BS - \delta - \epsilon}^{n(t)} 
\end{split}
\end{align}

for all times $t > 0$.

To show this, let us consider the perturbed system Eq.~\eqref{eq:system}. 
At each jump event $t_i$, the conditional probability to \emph{not} exit the basin of attraction 
is given by the shifted basin stability evaluated at the left limit $x_i$ of the 
trajectory before the jump:

\begin{align} 
\begin{split}
P\br{x(t_i^+) \in B \vert\, x(t_i^-) = x_i \in B} = \BS_{x_i}\;,
\end{split}
\end{align}

where $t_i^+$ and $t_i^-$ denote the right respectively left limit of $t$ to the jump time $t_i$.
Therefore, if we ensure that $\BS_{x_i}$ is close to $\BS$, 
we will also ensure that the perturbations are independent of each other in the sense we defined above.

Now given that the process is in $S$ before the jump at $t_i$, we want to understand what the probability 
is that it will return to $S$ before the next jump at $t_{i}$. If we started at the attractor rather 
than in $S$ this would be given by $\BS\br{\Delta t}$. The probability 
with respect to the shifted probability density thus differs from this at most by $\epsilon$. 
Assuming further that $\Delta t$ is larger than the independence time $T_{ind}$, Eq.~\eqref{eq:independence_time}
yields the lower bound:

\begin{align} 
\begin{split}
P\br{x(t_{i}^-) \in S\,\vert\, x(t_{i-1}^-) \in S} &\geq \BS\br{\Delta t} - \epsilon \\& \geq \BS - \delta - \epsilon 
\end{split}
\end{align}

Thus, for a sequence of consecutive jumps counted by $n\br{t}$, we find

\begin{align} 
\begin{split}
P_{remain}\br{t, x(0)} 
&\geq \prod\limits_{i=1}^{n(t)} P\br{x(t_{i}^-) \in S\,\vert\, x(t_{i-1}^-) \in S} \\
&\geq \prod\limits_{i=1}^{n(t)} \BS - \delta - \epsilon \\
&= \br{\BS - \delta - \epsilon }^{n(t)} \;.
\end{split}
\end{align}

The above formula applies as soon as the system enters the region bounded by $S$ once. 
Hence, if the stochastic process conditioned on staying in the basin of attraction has probability $1$ of hitting $S$, Eq.~\eqref{eq:remain-probability-epsilons} will also be the asymptotic form of the remain probability. Note also that the remain probability considers entire trajectories in the basin, not the probability to return there after having left.

The bound is neccesarily not tight, as it only considers trajectories that remain in the basin by returning to within $S$ before the next perturbation. We expect that for independence times corresponding to small $\delta$, this will be the dominant mechanism. For smaller times, there will be a non-negligible contribution to the remain probability from jumps that cancel each other out.

\section{A practical estimator} \label{sec:estimator}

The above arguments establish a lower bound for the remain probability, but they 
do not provide an effective way to evaluate the quantities involved. The main 
difficulty in constructing an efficient estimator lies in evaluating the metric 
$\In\br{x}$ and constructing a transverse return surface $S$ given an $\epsilon$. 
This problem simplifies considerably in the important special case that $\epsilon$ 
is chosen small enough that we only need to evaluate $\In\br{x}$ close to the attractor.
We now give an explicit formula based on the linearised dynamics for this case.

First let us consider $\In\br{x}$. We Taylor expand $\rho$ around the origin to first order, we find

\begin{align} \label{eq:in_approx}
\begin{split}
\In\br{x} \simeq \Vert x\Vert \int\limits_X \Vert \nabla \rho\br{x^\prime}\Vert dx^\prime = \frac{\Vert x\Vert}{C_\rho}\;,
\end{split}
\end{align}

defining a constant $C_\rho$ which is independent of the dynamics. It can be evaluated 
analytically for some common $\rho$, like uniform or Gaussian distributions, and numerically in general.

Thus all points inside the sphere $U_\epsilon = \{ x \vert\, \Vert\, x\Vert\, = \epsilon C_\rho \}$ 
satisfy $\In\br{x} \leq \epsilon$. This sphere might not be transverse, hence we are 
looking for a transverse surface $S_\epsilon$ of the time-tracking Lyapunov function entirely 
contained within $U_\epsilon$. This is schematically illustrated in Fig.~\ref{fig:schematic}b,
where the relation between $U_\epsilon$ and $S_\epsilon$ is indicated for a fictional 
multistable system with a fixed point $x^\ast$ and corresponding basin $B$.

As we are in a neighbourhood of the fixed point we can 
consider the linearised system associated to Eq.~\ref{eq:system_no_pert} given by

\begin{align}\label{eq:linear}
\begin{split}
\dot x(t) = J x(t) \;. 
\end{split}
\end{align}

If the Jacobian matrix $J$ is symmetric then $U_\epsilon$ is transverse, we can chose 
$S_\epsilon = U_\epsilon$ and are done.
To account for the general case, we can make use of quadratic Lyapunov functions 
$W\br{x} = x^\dagger  Lx$ for the linear system Eq.~\ref{eq:linear},
satisfying $\dot W\br{x} = x^\dagger Q x$ with $Q$ 
symmetric and negative definite. Given $J$ and a choice of $Q$, we can find 
a Lyapunov function by solving the matrix equation

\begin{align} 
\begin{split}
J^\dagger L + L J & = Q \;.
\end{split}
\end{align}

To find the maximum $\vert\, x\vert$ 
reached on the level set of $W\br{x}$, we differentiate $x^2$ in the direction parallel to the 
level set and look for extrema. Take a derivative $\partial_v = v \cdot \partial$. 
Then we require $\partial_v W\br{x} = 0$ for the derivative to be tangential to the level set. 
An extremum on the level set thus satisfies the following set of equations:

\begin{align} 
\begin{split}
\quad & \partial_v x^2 = 2 v^\dagger x = 0 \\
\forall v, \; \text{s.t.:} \quad & v^\dagger L x + x^\dagger L v = 2  v^\dagger L x  = 0
\end{split}
\end{align}

where we have used that $L$ is symmetric.
We immediately see that for $L = 1$, when our level sets are spheres, every point is an extremum.
In general, it follows that as $x$ is orthogonal to all $v$, and the $v$ span the space orthogonal to $Lx$, 
$x$ and $Lx$ need to be parallel. Thus, the extrema are in the eigendirections of $L$. 
The maximum $x_{max}^2$ for a given level set is achieved in the eigendirection to the smallest eigenvalue $\lambda$, thus the level set value is given by $W(x_{max}) = \lambda x_{max}^\dagger x_{max} =  \lambda (C_\rho \In\br{x_{max}})^2$.
The largest level set contained in $U_\epsilon$ is thus given by $x^\dagger L x = \lambda (\epsilon C_\rho)^2$.

Therefore, the transverse surface $S_\epsilon$ is defined as

\begin{align} \label{eq:return_set}
\begin{split}
S_\epsilon = \{ x \vert\, x^\dagger L x = \lambda (\epsilon C_\rho)^2 \} \; .
\end{split}
\end{align}

The fact that we have $L$ on the left and $\lambda$ on the right shows that this relation does not depend on an overall 
scaling factor of the Lyapunov function. To make $S_\epsilon$ as large as possible we want to make the ratios of the smallest eigenvalue of $L$ to the other ones, $\frac{\lambda_i}{\lambda}$, small. We leave the question, how to choose $Q$ such as to achieve this, open.

While direct Monte-Carlo estimation of finite-time basin stability with the specified $S_\epsilon$ will lead to a valid independence time, the surface chosen will typically be far from optimal. The optimal surface $S_\epsilon^\text{opt}$ can be defined by taking the surface $S_\epsilon$ and evolving every point on it backwards in time until its $\In$ distance to the attractor crosses $\epsilon$.

While this surface can not be constructed explicitly in general, if $\In(x)$ can be evaluated efficiently, we can evaluate the
finite-time basin stability with respect to $S_\epsilon^\text{opt}$ by backtracking along the trajectories. In practice this means we start by generating trajectories that run until they hit $S_\epsilon$, guaranteeing the $\In(x)$ will never grow larger than $\epsilon$ again at later times, and then backtracking along the trajectory to find the first time where $\In(x) > \epsilon$.

\section{A concrete example} \label{sec:results}

In the following, we demonstrate the effective estimator for independence times, as well as the main result on remain probabilities, in a benchmark dynamical system.


For higher dimensional systems evaluating the Lyapunov function explicitly is not feasible. However, a sampling-based approach, analogous to basin stability estimations (e.g. \cite{Menck2013}) can be applied here.

The Monte-Carlo sampling procedure is as follows: 

\begin{itemize}
\item Given a distribution $\rho$ and a tolerance $\epsilon$, determine $S_\epsilon$, 
		for instance using the method described in Sec.~\ref{sec:estimator}.
\item Sampling iteration:
	\begin{enumerate}
	\item Draw a random initial condition from $\rho_{x^\ast}$ centred at the fixed point.
	\item Integrate the unperturbed system (Eq.~\ref{eq:system_no_pert}) until either it reaches $S_\epsilon$
	or a cut-off time $T^c$ is reached. If it crosses $S_\epsilon$ record the time at which it does.
	\item (optional) Backtrack along the trajectory to record the time at which $\In(x)$ last crosses $\epsilon$
	\end{enumerate}
\end{itemize}

The sampling step should be repeated for a sufficient ensemble of initial conditions
to get significant statistics. Denote by $M_T$
the number of trajectories returning to $S_\epsilon$ within time $T$ or less and by
$N$ the total number of trajectories sampled.
Then, an estimator for the finite-time basin stability $\hat{\FTBS}\br{T}$ for $T < T^c$ is given by

\begin{align} \label{eq:ftbs_est}
\begin{split}
\hat{\FTBS}\br{T} = \frac{M_T}{N} 
\end{split}
\end{align}

with a standard error $e_{\hat{\FTBS}\br{T}}$ as

\begin{align} \label{eq:ftbs_est_error}
\begin{split}
e_{\hat{\FTBS}\br{T}} = \sqrt{\frac{\hat{\FTBS}\br{T} \br{1 - \hat{\FTBS}\br{T}}}{N}} \; ,
\end{split}
\end{align}

since for a fixed $T$ we can regard this as a Bernoulli experiment, because trajectories either return or not.
Note that if $\hat{\FTBS}\br{T}\approx 1$ or $\hat{\FTBS}\br{T}\approx 0$, more robust estimators are 
available \cite{Agresti1998}.

Note that while the error decreases with the number of samples and does not depend on the dimensionality of the system, the time taken to evaluate a sample does depend on the system dimension at least linearly.

We will illustrate this by using the damped-driven pendulum as a benchmark system:

\begin{align} \label{eq:swing}
\begin{split}
\dot\phi &= \omega\\
\dot\omega &= p - \alpha\omega - k \sin\br{\phi + \arcsin\frac{p}{k}}
\end{split}
\end{align}

with $p=1$, $\alpha=0.1$ and $k=8$. For this set of parameters, the system has
two attractors, namely a limit cycle and a fixed point 
$x^\ast = \br{\phi^\ast,\omega^\ast} = \br{0, 0}$ at the origin\footnote{Note that we applied a 
phase shift of $\arcsin\br{p/k}$ to set the fix point to the origin.}.

For illustrative purposes, we choose a distribution $\rho\br{x}$ 
to draw uniformly distributed perturbations at a point $x=\br{\phi, \omega}$ 
from the box $R\br{\phi, \omega} =\left[\phi-\pi/3;\phi + \pi/3\right]\times\left[\omega -5;\omega +5\right]$.
This way, $R$ is almost entirely overlapping with the bulk of the basin of attraction 
of $x^\ast$ (cf. Fig.~\ref{fig:schematic}a for a schematic), such that we can expect 
$\BS$ to be close to $1$. Still, as we will see below, $\FTBS\br{T}$ can deviate strongly 
from $\BS$, especially for small $T$. 	

To ensure sufficient statistics, we use a sample size of $N=20,000$ points.

\begin{figure}[!t]
    \centering
    \begin{subfigure}[t]{0.45\textwidth}
        \centering
        \includegraphics[width=0.97\columnwidth]{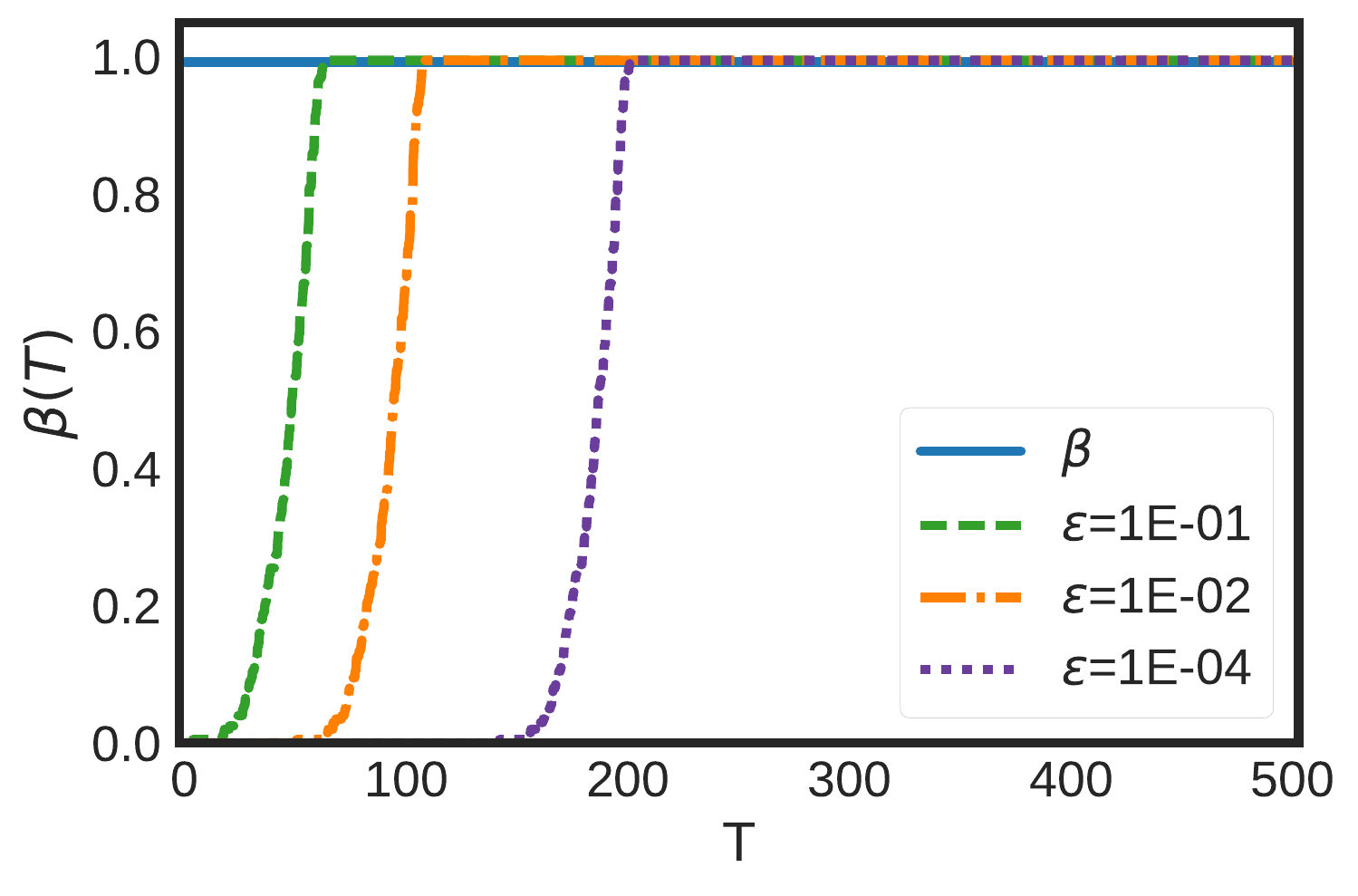}
        \caption{}
    \end{subfigure}%
    
    \begin{subfigure}[t]{0.45\textwidth}
        \centering
        \includegraphics[width=0.97\columnwidth]{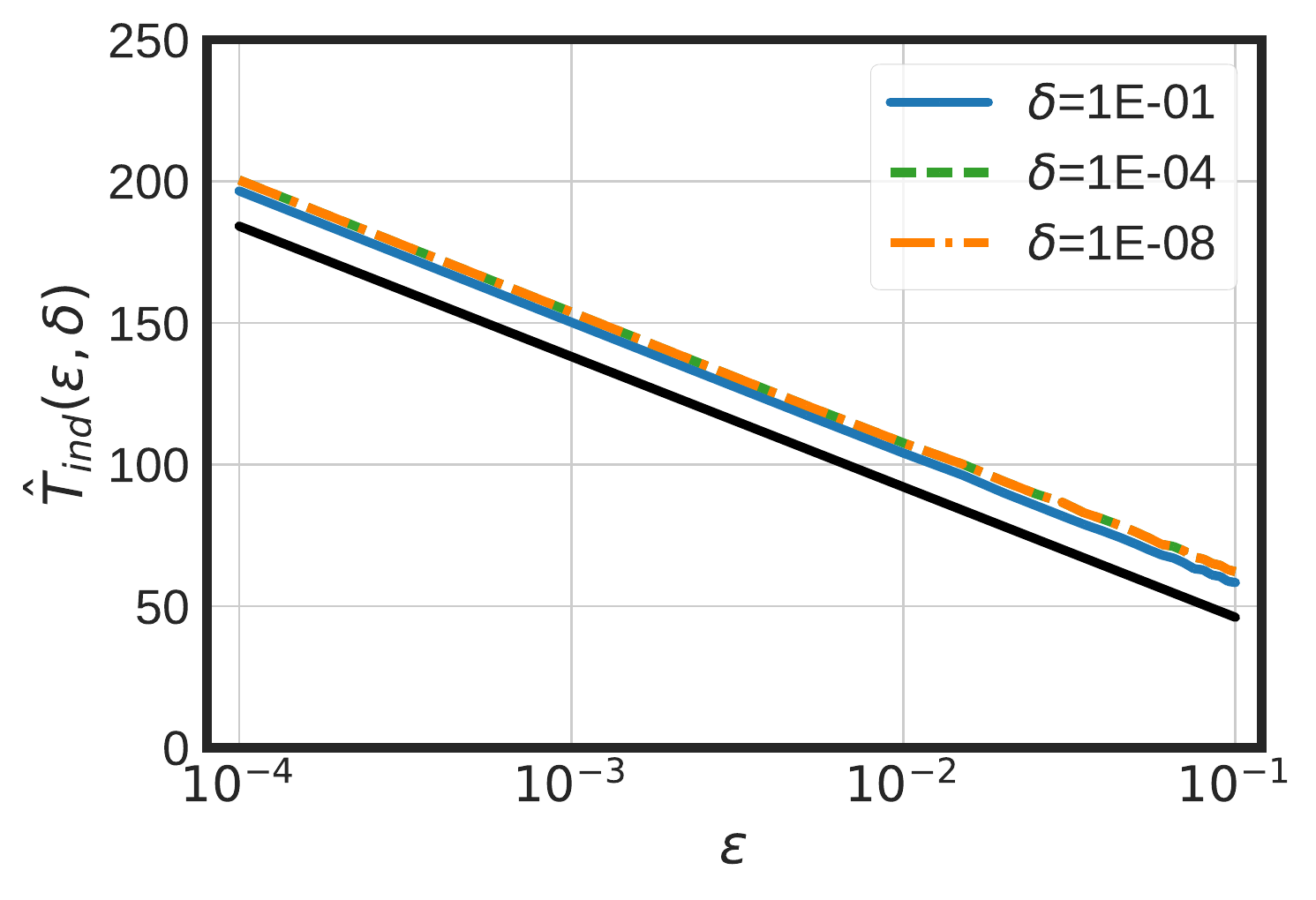}
        \caption{}
    \end{subfigure}
    \caption{
    \textbf{(a)} 
    The finite time basin stability $\FTBS$ curve (Eq.~\ref{eq:ftbs}) for various $\epsilon$. 
    The solid blue line at the top gives the basin stability (Eq.~\ref{eq:bs}).
    \textbf{(b)} 
    The independence time $T_{ind}$ for various $\delta$ as a function of $\epsilon$.    
    }
   \label{fig:ftbs_curve}
\end{figure}

For this specific choice of $\rho$, we determine $S_\epsilon$ using Eq.~\ref{eq:return_set} to be

\begin{equation}
S_\epsilon = \{ x \vert\, x^\dagger L' x = \epsilon^2 \} \; ,
\end{equation}

where $L' = \frac{L}{\lambda C_{\rho}^2}$ is given by

\begin{equation}
L' = 
\begin{pmatrix}
5.95152498 & 0.00838866 \\
0.00838866 &  0.74971598
\end{pmatrix}\; .
\end{equation}

Fig.~\ref{fig:ftbs_curve}a summarises the results for the system Eq.~\ref{eq:swing}. The horizontal
blue line denotes the basin stability estimation $\hat\beta=0.9874$ which is close to $1$ as expected due to our choice 
of $R$. Indeed, beyond a certain time scale that depends on $\epsilon$, we observe that the finite-time basin 
stability curves approach the value of $\BS$. From these points, we estimate the independence times
$\hat{T}_{ind}\br{\epsilon,\delta}$ depicted in Fig.~\ref{fig:ftbs_curve}b using Eq.~\ref{eq:independence_time}.
As indicated by Fig.~\ref{fig:ftbs_curve}b, our results suggest that there is no significant dependence
on the tolerance parameter $\delta$ for this particular system. Apparently, there is a rather sudden transition
towards the value of $\hat{\BS}$ that cannot be resolved by the numerical differences of $T$-values.
The crucial parameter here is $\epsilon$ determining the extent of the return set $S_\epsilon$. 
The logarithmic scale in Fig.~\ref{fig:ftbs_curve}b underlines that the independence time depends
exponentially on the tolerance $\epsilon$ as the corresponding $S_\epsilon$ encloses the 
asymptotically stable fixed point $x^\ast$ ever closer. Hence, the scaling 
$\hat{T}_{ind}\br{\epsilon,\delta} \sim \exp{\lambda\epsilon}$ seems to be determined by the
real part $\lambda=-0.05$ of the two conjugate Jacobian eigenvalues of Eq.~\ref{eq:swing}
linearised at $x^\ast$. This is indicated by the solid black line in Fig.~\ref{fig:ftbs_curve}b
which has a slope of $\lambda$.

\begin{figure}[!t]
    \centering
    \begin{subfigure}[t]{0.45\textwidth}
        \centering
        \includegraphics[width=0.95\columnwidth]{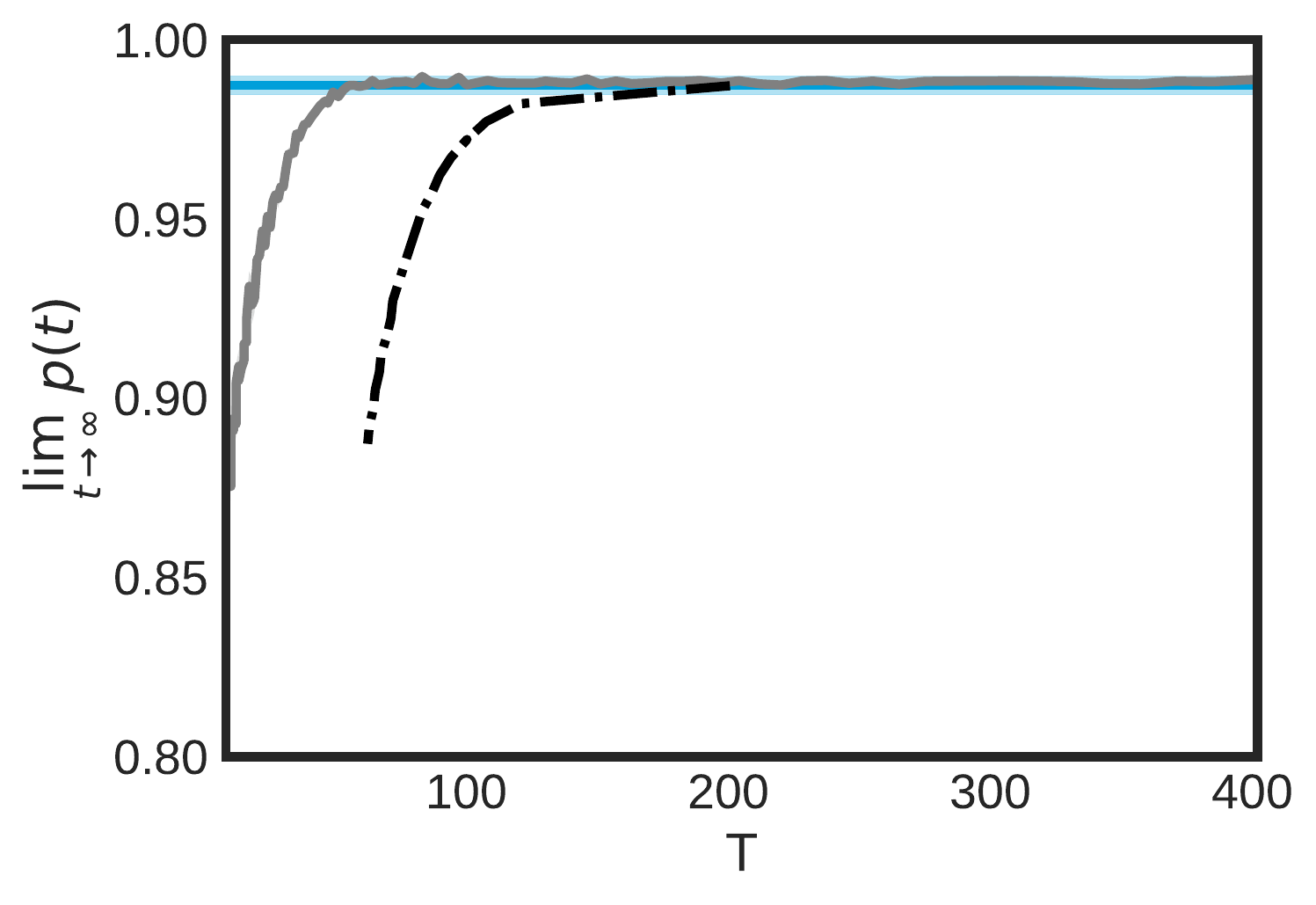}
        \caption{}
    \end{subfigure}%
    
    \begin{subfigure}[t]{0.45\textwidth}
        \centering
        \includegraphics[width=0.9\columnwidth]{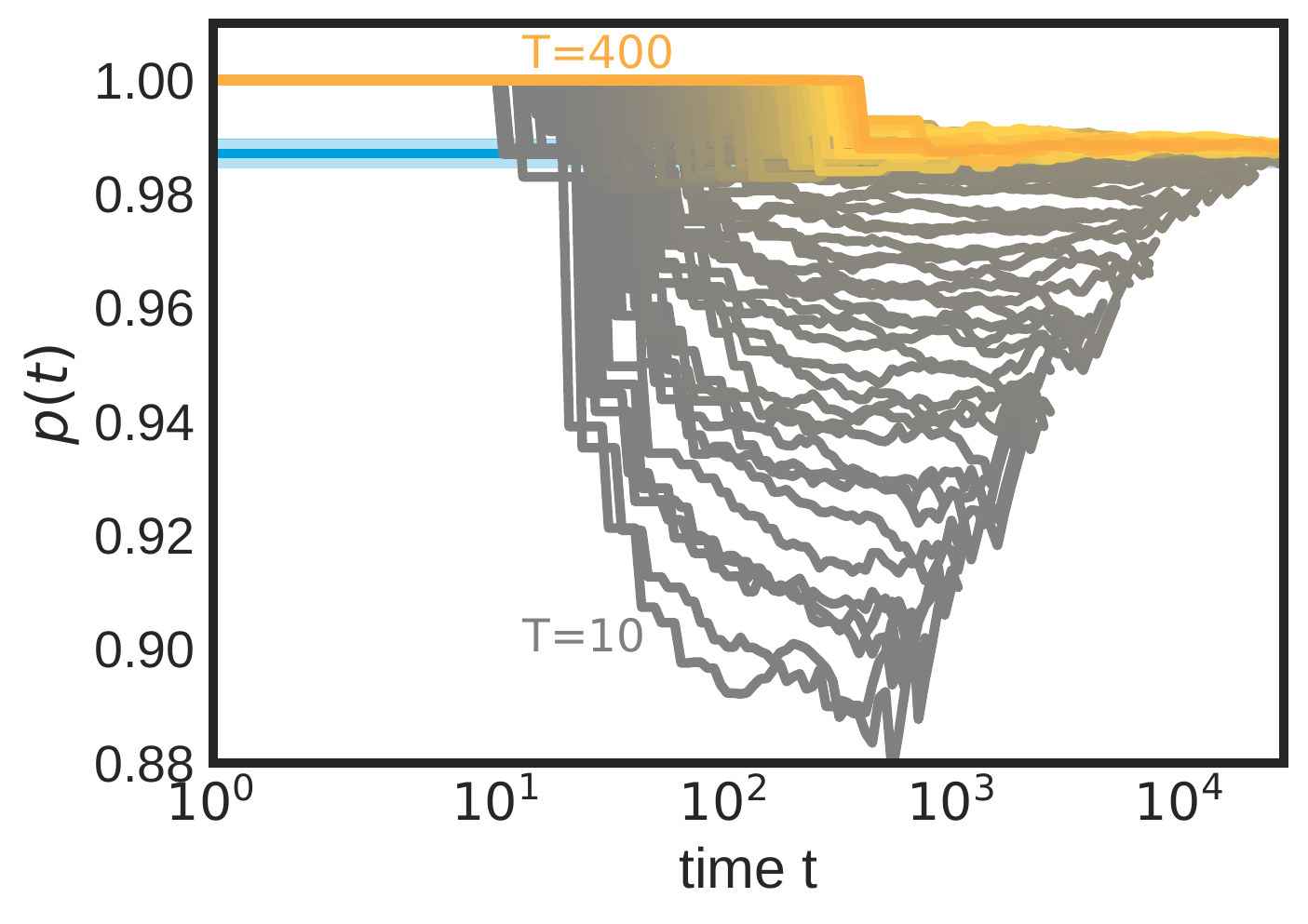}
        \caption{}
    \end{subfigure} 
    \caption{ 
    In both figures, the horizontal blue line shows the estimate for the basin
	stability, the shading indicates one standard error.
    \textbf{(a)} 
	The solid line is the numerically determined asymptotic
	remain probability per jump event, $\lim_{t \rightarrow \infty}p(t)$ for perturbations
	that are at least $T$ apart, and the dashed dotted line shows the lower bound $\beta - \epsilon(T_{ind}) - \delta$ 
	associated to an independence time $T_{ind}$, with $\delta = 10^{-8}$.
	The function $\epsilon(T_{ind})$ is the inverse of the estimated
	$\hat{T}_{ind}\br{\epsilon, \delta}$ at fixed $\delta$.
	As therethe dependence on $\delta$ is negligable (Fig.~\ref{fig:ftbs_curve}b), we only depict the bound for $\delta=10^{-8}$.
	\textbf{(b)} The
	picture shows the remain probability $p(t)$ as a function of time $t$ given a
	particular perturbation interval $T$. Each curve corresponds to the
	fraction of trajectories that remain in the basin, perturbed at an interval $T$
	which is indicated by the color progressing from grey to orange. 
    }
   \label{fig:p_remain}
\end{figure}

We can now illustrate our result Eq.~\ref{eq:remain-probability-epsilons} for the probability 
to remain within the basin of attraction up to a certain time, given that we start in  $S_\epsilon$ near 
the origin. For this, we simulate an ensemble of random processes by adding a jump process to the dynamics
Eq.~\ref{eq:swing} (cf. Eq.~\ref{eq:system}). Explicitly, we choose $100$ different time intervals $T$ 
between $10$ and $400$ time units such that after each interval a deviation is randomly selected according
 to a uniform distribution $\rho_{x^\prime}$ centred at the current state $x^\prime$ with a shifted domain $R$ as above. 
For each choice of $T$ we estimate the escape time distribution by recording the first time a trajectory 
jumps outside the origin's basin of attraction using an ensemble size of $N=1000$ trajectories. Denoting the
number of trajectories with an escape time larger than $t$ by $N_{>}\br{t}$, we estimate the remain 
probability as $\hat{P}_{remain} = N_{>}\br{t} / N$. Then rewriting Eq.~\ref{eq:remain-probability-epsilons} as a per jump probability
yields the following relation

\begin{equation}
p\br{t} = \hat{P}_{remain}^{\frac{1}{n\br{t}}} \geq \hat{\BS} - \epsilon - \delta \;,
\end{equation}

which we expect to hold if $T$ is larger than the corresponding independence time. 
For the system Eq.~\ref{eq:swing}, the independence time for $\epsilon = 10^{-1}$ is given by $\approx 60$, for $\epsilon = 10^{-2}$ it is $\approx 100$. We see in Fig.~\ref{fig:p_remain} that perturbations spaced $60$ apart can not destabilise the system at a rate greater than ($\BS - 0.1$), and after $T = 100$ we are within $0.01$ of the basin stability asymptotic estimate (and thus close to its sampling error), as predicted. Further, by plotting the lower bound (for fixed $\delta = 10^{-8}$) as a function of the independence time it is associated to, we see that our bound is satisfied across all times.

\section{Discussion} \label{sec:discussion}

Just as for asymptotic basin stability, finite-time basin stability admits a simple and efficient sampling-based estimator 
that works for systems with a high number of dimensions.
If the asymptotic basin stability is equal to one, this allows us to effectively guarantee, up to specified errors, that perturbations that occur at least the independence time apart can not destabilise a system.
We expect there to be a wide array of applications to the question, how rare large events have to be to not destabilise the system, which we intend to explore in future work.

We have also seen that the lower bound for which we developed the estimator is not sharp. This is entirely due to the estimate in Eq.~\eqref{eq:iota estimate}, which bounds the shifted basin stability through the distance measure $\iota$. One challenge for future work is to develop and prove an effective estimator that can sidestep the use of $\iota$, and directly assess the escape probability.

More generally we see under which conditions basin stability can be seen as the remain probability in the basin of attraction for systems subject to rare, strong events. Given the frequency of perturbations, basin stability completely determines the escape rate from the basin in this case.

One interesting analogue to our work is the study of the exit time distribution for basin escapes in systems subject to Levy noise \cite{Serdukova2016Stochastic,Serdukova2017Metastability}.
The type of stochastic process studied here, deterministic with interspersed jumps, can be used to approximate such L\'{e}vy processes in some asymptotic regime \cite{Imkeller2006,Pavlyukevich2007,Pavlyukevich2007a}. We expect that the results of this paper can be used to develop estimators that can quantify when this asymptotic regime is reached. Consequently, it should lead to more efficient ways to perform
an analysis as in \cite{Serdukova2017Metastability}. 

An open question for future work is to extend the notions discussed here to non-fixed point attractors. The main challenge here will lie in building a practical estimator that works.

\section*{Acknowledgements}

PS, FH and JK acknowledge the support of BMBF, CoNDyNet, FK. 03SF0472A. KW was supported by the European Comissions Marie Curie Fellowship (Grant No. 660616). This work was supported by the Volkswagen Foundation 
(Grant No. 88462). Funded by the Deutsche Forschungsgemeinschaft (DFG, German Research Foundation) – KU 837/39-1 / RA 516/13-1.
All authors gratefully acknowledge the European Regional Development Fund (ERDF), the German Federal Ministry of Education and Research and the Land Brandenburg for supporting this project by providing resources on the high performance computer system at the Potsdam Institute for Climate Impact Research.


%
%
%
%
%
%
%

\bibliography{lib}

\begin{thebibliography}{29}%
\makeatletter
\providecommand \@ifxundefined [1]{%
 \@ifx{#1\undefined}
}%
\providecommand \@ifnum [1]{%
 \ifnum #1\expandafter \@firstoftwo
 \else \expandafter \@secondoftwo
 \fi
}%
\providecommand \@ifx [1]{%
 \ifx #1\expandafter \@firstoftwo
 \else \expandafter \@secondoftwo
 \fi
}%
\providecommand \natexlab [1]{#1}%
\providecommand \enquote  [1]{``#1''}%
\providecommand \bibnamefont  [1]{#1}%
\providecommand \bibfnamefont [1]{#1}%
\providecommand \citenamefont [1]{#1}%
\providecommand \href@noop [0]{\@secondoftwo}%
\providecommand \href [0]{\begingroup \@sanitize@url \@href}%
\providecommand \@href[1]{\@@startlink{#1}\@@href}%
\providecommand \@@href[1]{\endgroup#1\@@endlink}%
\providecommand \@sanitize@url [0]{\catcode `\\12\catcode `\$12\catcode
  `\&12\catcode `\#12\catcode `\^12\catcode `\_12\catcode `\%12\relax}%
\providecommand \@@startlink[1]{}%
\providecommand \@@endlink[0]{}%
\providecommand \url  [0]{\begingroup\@sanitize@url \@url }%
\providecommand \@url [1]{\endgroup\@href {#1}{\urlprefix }}%
\providecommand \urlprefix  [0]{URL }%
\providecommand \Eprint [0]{\href }%
\providecommand \doibase [0]{http://dx.doi.org/}%
\providecommand \selectlanguage [0]{\@gobble}%
\providecommand \bibinfo  [0]{\@secondoftwo}%
\providecommand \bibfield  [0]{\@secondoftwo}%
\providecommand \translation [1]{[#1]}%
\providecommand \BibitemOpen [0]{}%
\providecommand \bibitemStop [0]{}%
\providecommand \bibitemNoStop [0]{.\EOS\space}%
\providecommand \EOS [0]{\spacefactor3000\relax}%
\providecommand \BibitemShut  [1]{\csname bibitem#1\endcsname}%
\let\auto@bib@innerbib\@empty
\bibitem [{\citenamefont {Lyapunov}(1907)}]{Lyapunov1907}%
  \BibitemOpen
  \bibfield  {author} {\bibinfo {author} {\bibfnamefont {A.~M.}\ \bibnamefont
  {Lyapunov}},\ }\bibfield  {title} {\enquote {\bibinfo {title} {{Probl\`{e}me
  G\'{e}n\'{e}ral de la Stabilit\'{e} du Mouvement}},}\ }\href@noop {}
  {\bibfield  {journal} {\bibinfo  {journal} {Annales de la Facult{\'{e}} des
  sciences de Toulouse: Math{\'{e}}matiques}\ }\textbf {\bibinfo {volume}
  {2}},\ \bibinfo {pages} {203--474} (\bibinfo {year} {1907})}\BibitemShut
  {NoStop}%
\bibitem [{\citenamefont {Wiley}, \citenamefont {Strogatz},\ and\ \citenamefont
  {Girvan}(2006)}]{Wiley2006}%
  \BibitemOpen
  \bibfield  {author} {\bibinfo {author} {\bibfnamefont {D.~A.}\ \bibnamefont
  {Wiley}}, \bibinfo {author} {\bibfnamefont {S.~H.}\ \bibnamefont {Strogatz}},
  \ and\ \bibinfo {author} {\bibfnamefont {M.}~\bibnamefont {Girvan}},\
  }\bibfield  {title} {\enquote {\bibinfo {title} {{The size of the sync
  basin}},}\ }\href@noop {} {\bibfield  {journal} {\bibinfo  {journal} {Chaos:
  An Interdisciplinary Journal of Nonlinear Science}\ }\textbf {\bibinfo
  {volume} {16}},\ \bibinfo {pages} {015103} (\bibinfo {year}
  {2006})}\BibitemShut {NoStop}%
\bibitem [{\citenamefont {Klinshov}, \citenamefont {Nekorkin},\ and\
  \citenamefont {Kurths}(2015)}]{Klinshov2015}%
  \BibitemOpen
  \bibfield  {author} {\bibinfo {author} {\bibfnamefont {V.~V.}\ \bibnamefont
  {Klinshov}}, \bibinfo {author} {\bibfnamefont {V.~I.}\ \bibnamefont
  {Nekorkin}}, \ and\ \bibinfo {author} {\bibfnamefont {J.}~\bibnamefont
  {Kurths}},\ }\bibfield  {title} {\enquote {\bibinfo {title} {{Stability
  threshold approach for complex dynamical systems}},}\ }\href@noop {}
  {\bibfield  {journal} {\bibinfo  {journal} {New J. Phys.}\ }\textbf {\bibinfo
  {volume} {18}},\ \bibinfo {pages} {013004} (\bibinfo {year}
  {2015})}\BibitemShut {NoStop}%
\bibitem [{\citenamefont {Mitra}, \citenamefont {Kurths},\ and\ \citenamefont
  {Donner}(2015)}]{Mitra2015}%
  \BibitemOpen
  \bibfield  {author} {\bibinfo {author} {\bibfnamefont {C.}~\bibnamefont
  {Mitra}}, \bibinfo {author} {\bibfnamefont {J.}~\bibnamefont {Kurths}}, \
  and\ \bibinfo {author} {\bibfnamefont {R.~V.}\ \bibnamefont {Donner}},\
  }\bibfield  {title} {\enquote {\bibinfo {title} {{An integrative quantifier
  of multistability in complex systems based on ecological resilience}},}\
  }\href@noop {} {\bibfield  {journal} {\bibinfo  {journal} {Sci. Rep.}\
  }\textbf {\bibinfo {volume} {5}},\ \bibinfo {pages} {16196} (\bibinfo {year}
  {2015})}\BibitemShut {NoStop}%
\bibitem [{\citenamefont {Hahn}(1958)}]{Hahn1958}%
  \BibitemOpen
  \bibfield  {author} {\bibinfo {author} {\bibfnamefont {W.}~\bibnamefont
  {Hahn}},\ }\bibfield  {title} {\enquote {\bibinfo {title} {{{\"{U}}ber die
  Anwendung der Methode von Ljapunov auf Differenzengleichungen}},}\
  }\href@noop {} {\bibfield  {journal} {\bibinfo  {journal} {Mathematische
  Annalen}\ }\textbf {\bibinfo {volume} {136}},\ \bibinfo {pages} {430--441}
  (\bibinfo {year} {1958})}\BibitemShut {NoStop}%
\bibitem [{\citenamefont {Malisoff}\ and\ \citenamefont
  {Mazenc}(2009)}]{Malisoff2009}%
  \BibitemOpen
  \bibfield  {author} {\bibinfo {author} {\bibfnamefont {M.}~\bibnamefont
  {Malisoff}}\ and\ \bibinfo {author} {\bibfnamefont {F.}~\bibnamefont
  {Mazenc}},\ }\href@noop {} {\emph {\bibinfo {title} {{Constructions of Strict
  {Lyapunov} Functions}}}},\ \bibinfo {edition} {1st}\ ed.,\ Communications and
  Control Engineering\ (\bibinfo  {publisher} {Springer London},\ \bibinfo
  {address} {London},\ \bibinfo {year} {2009})\ pp.\ \bibinfo {pages} {XVI,
  386}\BibitemShut {NoStop}%
\bibitem [{\citenamefont {Giesl}\ and\ \citenamefont
  {Hafstein}(2015)}]{Giesl2015}%
  \BibitemOpen
  \bibfield  {author} {\bibinfo {author} {\bibfnamefont {P.}~\bibnamefont
  {Giesl}}\ and\ \bibinfo {author} {\bibfnamefont {S.}~\bibnamefont
  {Hafstein}},\ }\bibfield  {title} {\enquote {\bibinfo {title} {{Review on
  computational methods for {Lyapunov} functions}},}\ }\href@noop {} {\bibfield
   {journal} {\bibinfo  {journal} {Discrete Contin. Dyn. Syst. Ser. B}\
  }\textbf {\bibinfo {volume} {20}},\ \bibinfo {pages} {2291--2331} (\bibinfo
  {year} {2015})}\BibitemShut {NoStop}%
\bibitem [{\citenamefont {Graham}\ and\ \citenamefont
  {T{\'{e}}l}(1984)}]{Graham1984}%
  \BibitemOpen
  \bibfield  {author} {\bibinfo {author} {\bibfnamefont {R.}~\bibnamefont
  {Graham}}\ and\ \bibinfo {author} {\bibfnamefont {T.}~\bibnamefont
  {T{\'{e}}l}},\ }\bibfield  {title} {\enquote {\bibinfo {title} {{Existence of
  a Potential for Dissipative Dynamical Systems}},}\ }\href@noop {} {\bibfield
  {journal} {\bibinfo  {journal} {Phys. Rev. Lett.}\ }\textbf {\bibinfo
  {volume} {52}},\ \bibinfo {pages} {9--12} (\bibinfo {year}
  {1984})}\BibitemShut {NoStop}%
\bibitem [{\citenamefont {Graham}, \citenamefont {Hamm},\ and\ \citenamefont
  {T{\'{e}}l}(1991)}]{Graham1991}%
  \BibitemOpen
  \bibfield  {author} {\bibinfo {author} {\bibfnamefont {R.}~\bibnamefont
  {Graham}}, \bibinfo {author} {\bibfnamefont {A.}~\bibnamefont {Hamm}}, \ and\
  \bibinfo {author} {\bibfnamefont {T.}~\bibnamefont {T{\'{e}}l}},\ }\bibfield
  {title} {\enquote {\bibinfo {title} {{Nonequilibrium potentials for dynamical
  systems with fractal attractors or repellers}},}\ }\href@noop {} {\bibfield
  {journal} {\bibinfo  {journal} {Phys. Rev. Lett.}\ }\textbf {\bibinfo
  {volume} {66}},\ \bibinfo {pages} {3089--3092} (\bibinfo {year}
  {1991})}\BibitemShut {NoStop}%
\bibitem [{\citenamefont {Parrilo}(2000)}]{Parrilo2000}%
  \BibitemOpen
  \bibfield  {author} {\bibinfo {author} {\bibfnamefont {P.}~\bibnamefont
  {Parrilo}},\ }\emph {\bibinfo {title} {{Structured Semidefinite Programs and
  Semialgebraic Geometry Methods in Robustness and Optimiziation}}},\
  \href@noop {} {Ph.D. thesis},\ \bibinfo  {school} {California Institute of
  Technology, Pasadena, CA} (\bibinfo {year} {2000})\BibitemShut {NoStop}%
\bibitem [{\citenamefont {Hafstein}(2004)}]{Hafstein2004}%
  \BibitemOpen
  \bibfield  {author} {\bibinfo {author} {\bibfnamefont {S.}~\bibnamefont
  {Hafstein}},\ }\bibfield  {title} {\enquote {\bibinfo {title} {{A
  constructive converse {Lyapunov} theorem on exponential stability}},}\
  }\href@noop {} {\bibfield  {journal} {\bibinfo  {journal} {Discrete Contin.
  Dyn. Syst.}\ }\textbf {\bibinfo {volume} {10}},\ \bibinfo {pages} {657--678}
  (\bibinfo {year} {2004})}\BibitemShut {NoStop}%
\bibitem [{\citenamefont {Giesl}(2007)}]{Giesl2007}%
  \BibitemOpen
  \bibfield  {author} {\bibinfo {author} {\bibfnamefont {P.}~\bibnamefont
  {Giesl}},\ }\href@noop {} {\emph {\bibinfo {title} {Construction of Global
  {Lyapunov} Functions using Radial Basis Functions}}},\ \bibinfo {series}
  {Lecture Notes in Math.}, Vol.\ \bibinfo {volume} {1904}\ (\bibinfo
  {publisher} {Springer, Berlin},\ \bibinfo {year} {2007})\BibitemShut
  {NoStop}%
\bibitem [{\citenamefont {Camilli}, \citenamefont {Gr\"{u}ne},\ and\
  \citenamefont {Wirth}(2001)}]{Camilli2001}%
  \BibitemOpen
  \bibfield  {author} {\bibinfo {author} {\bibfnamefont {F.}~\bibnamefont
  {Camilli}}, \bibinfo {author} {\bibfnamefont {L.}~\bibnamefont {Gr\"{u}ne}},
  \ and\ \bibinfo {author} {\bibfnamefont {F.}~\bibnamefont {Wirth}},\
  }\bibfield  {title} {\enquote {\bibinfo {title} {{A generalization of Zubov's
  method to perturbed systems}},}\ }\href@noop {} {\bibfield  {journal}
  {\bibinfo  {journal} {SIAM J. Control Optim.}\ }\textbf {\bibinfo {volume}
  {40}},\ \bibinfo {pages} {496--515} (\bibinfo {year} {2001})}\BibitemShut
  {NoStop}%
\bibitem [{\citenamefont {Chiang}(2010)}]{Chiang2010}%
  \BibitemOpen
  \bibfield  {author} {\bibinfo {author} {\bibfnamefont {H.-D.}\ \bibnamefont
  {Chiang}},\ }\href@noop {} {\emph {\bibinfo {title} {{Direct Methods for
  Stability Analysis of Electric Power Systems}}}}\ (\bibinfo  {publisher}
  {John Wiley {\&} Sons, Inc.},\ \bibinfo {address} {Hoboken, NJ, USA},\
  \bibinfo {year} {2010})\BibitemShut {NoStop}%
\bibitem [{\citenamefont {Gajduk}, \citenamefont {Todorovski},\ and\
  \citenamefont {Kocarev}(2014)}]{Gajduk2014}%
  \BibitemOpen
  \bibfield  {author} {\bibinfo {author} {\bibfnamefont {A.}~\bibnamefont
  {Gajduk}}, \bibinfo {author} {\bibfnamefont {M.}~\bibnamefont {Todorovski}},
  \ and\ \bibinfo {author} {\bibfnamefont {L.}~\bibnamefont {Kocarev}},\
  }\bibfield  {title} {\enquote {\bibinfo {title} {{Stability of power grids:
  An overview}},}\ }\href@noop {} {\bibfield  {journal} {\bibinfo  {journal}
  {The European Physical Journal: Special Topics}\ }\textbf {\bibinfo {volume}
  {223}},\ \bibinfo {pages} {2387--2409} (\bibinfo {year} {2014})}\BibitemShut
  {NoStop}%
\bibitem [{\citenamefont {Menck}\ \emph {et~al.}(2013)\citenamefont {Menck},
  \citenamefont {Heitzig}, \citenamefont {Marwan},\ and\ \citenamefont
  {Kurths}}]{Menck2013}%
  \BibitemOpen
  \bibfield  {author} {\bibinfo {author} {\bibfnamefont {P.~J.}\ \bibnamefont
  {Menck}}, \bibinfo {author} {\bibfnamefont {J.}~\bibnamefont {Heitzig}},
  \bibinfo {author} {\bibfnamefont {N.}~\bibnamefont {Marwan}}, \ and\ \bibinfo
  {author} {\bibfnamefont {J.}~\bibnamefont {Kurths}},\ }\bibfield  {title}
  {\enquote {\bibinfo {title} {How basin stability complements the
  linear-stability paradigm},}\ }\href@noop {} {\bibfield  {journal} {\bibinfo
  {journal} {Nature Physics}\ }\textbf {\bibinfo {volume} {9}},\ \bibinfo
  {pages} {89--92} (\bibinfo {year} {2013})}\BibitemShut {NoStop}%
\bibitem [{\citenamefont {Schultz}\ \emph {et~al.}(2017)\citenamefont
  {Schultz}, \citenamefont {Menck}, \citenamefont {Heitzig},\ and\
  \citenamefont {Kurths}}]{Schultz2017}%
  \BibitemOpen
  \bibfield  {author} {\bibinfo {author} {\bibfnamefont {P.}~\bibnamefont
  {Schultz}}, \bibinfo {author} {\bibfnamefont {P.~J.}\ \bibnamefont {Menck}},
  \bibinfo {author} {\bibfnamefont {J.}~\bibnamefont {Heitzig}}, \ and\
  \bibinfo {author} {\bibfnamefont {J.}~\bibnamefont {Kurths}},\ }\bibfield
  {title} {\enquote {\bibinfo {title} {{Potentials and limits to basin
  stability estimation}},}\ }\href@noop {} {\bibfield  {journal} {\bibinfo
  {journal} {New Journal of Physics}\ }\textbf {\bibinfo {volume} {19}},\
  \bibinfo {pages} {023005} (\bibinfo {year} {2017})}\BibitemShut {NoStop}%
\bibitem [{\citenamefont {Mitra}\ \emph
  {et~al.}(2017{\natexlab{a}})\citenamefont {Mitra}, \citenamefont {Choudhary},
  \citenamefont {Sinha}, \citenamefont {Kurths},\ and\ \citenamefont
  {Donner}}]{Mitra2017}%
  \BibitemOpen
  \bibfield  {author} {\bibinfo {author} {\bibfnamefont {C.}~\bibnamefont
  {Mitra}}, \bibinfo {author} {\bibfnamefont {A.}~\bibnamefont {Choudhary}},
  \bibinfo {author} {\bibfnamefont {S.}~\bibnamefont {Sinha}}, \bibinfo
  {author} {\bibfnamefont {J.}~\bibnamefont {Kurths}}, \ and\ \bibinfo {author}
  {\bibfnamefont {R.~V.}\ \bibnamefont {Donner}},\ }\bibfield  {title}
  {\enquote {\bibinfo {title} {{Multiple-node basin stability in complex
  dynamical networks}},}\ }\href@noop {} {\bibfield  {journal} {\bibinfo
  {journal} {Phys. Rev. E}\ }\textbf {\bibinfo {volume} {95}},\ \bibinfo
  {pages} {032317} (\bibinfo {year} {2017}{\natexlab{a}})}\BibitemShut
  {NoStop}%
\bibitem [{\citenamefont {Rega}\ and\ \citenamefont {Lenci}(2005)}]{Rega2005}%
  \BibitemOpen
  \bibfield  {author} {\bibinfo {author} {\bibfnamefont {G.}~\bibnamefont
  {Rega}}\ and\ \bibinfo {author} {\bibfnamefont {S.}~\bibnamefont {Lenci}},\
  }\bibfield  {title} {\enquote {\bibinfo {title} {{Identifying, evaluating,
  and controlling dynamical integrity measures in non-linear mechanical
  oscillators}},}\ }\href@noop {} {\bibfield  {journal} {\bibinfo  {journal}
  {Nonlinear Analysis: Theory, Methods and Applications}\ }\textbf {\bibinfo
  {volume} {63}},\ \bibinfo {pages} {902--914} (\bibinfo {year}
  {2005})}\BibitemShut {NoStop}%
\bibitem [{\citenamefont {Hellmann}\ \emph {et~al.}(2016)\citenamefont
  {Hellmann}, \citenamefont {Schultz}, \citenamefont {Grabow}, \citenamefont
  {Heitzig},\ and\ \citenamefont {Kurths}}]{Hellmann2016}%
  \BibitemOpen
  \bibfield  {author} {\bibinfo {author} {\bibfnamefont {F.}~\bibnamefont
  {Hellmann}}, \bibinfo {author} {\bibfnamefont {P.}~\bibnamefont {Schultz}},
  \bibinfo {author} {\bibfnamefont {C.}~\bibnamefont {Grabow}}, \bibinfo
  {author} {\bibfnamefont {J.}~\bibnamefont {Heitzig}}, \ and\ \bibinfo
  {author} {\bibfnamefont {J.}~\bibnamefont {Kurths}},\ }\bibfield  {title}
  {\enquote {\bibinfo {title} {{Survivability of Deterministic Dynamical
  Systems}},}\ }\href@noop {} {\bibfield  {journal} {\bibinfo  {journal} {Sci.
  Rep.}\ }\textbf {\bibinfo {volume} {6}},\ \bibinfo {pages} {29654} (\bibinfo
  {year} {2016})}\BibitemShut {NoStop}%
\bibitem [{\citenamefont {Kittel}\ \emph {et~al.}(2017)\citenamefont {Kittel},
  \citenamefont {Heitzig}, \citenamefont {Webster},\ and\ \citenamefont
  {Kurths}}]{Kittel2016}%
  \BibitemOpen
  \bibfield  {author} {\bibinfo {author} {\bibfnamefont {T.}~\bibnamefont
  {Kittel}}, \bibinfo {author} {\bibfnamefont {J.}~\bibnamefont {Heitzig}},
  \bibinfo {author} {\bibfnamefont {K.}~\bibnamefont {Webster}}, \ and\
  \bibinfo {author} {\bibfnamefont {J.}~\bibnamefont {Kurths}},\ }\bibfield
  {title} {\enquote {\bibinfo {title} {{Timing of transients: quantifying
  reaching times and transient behavior in complex systems}},}\ }\href@noop {}
  {\bibfield  {journal} {\bibinfo  {journal} {New J. Phys.}\ }\textbf {\bibinfo
  {volume} {19}},\ \bibinfo {pages} {083005} (\bibinfo {year}
  {2017})}\BibitemShut {NoStop}%
\bibitem [{\citenamefont {Mitra}\ \emph
  {et~al.}(2017{\natexlab{b}})\citenamefont {Mitra}, \citenamefont {Kittel},
  \citenamefont {Choudhary}, \citenamefont {Kurths},\ and\ \citenamefont
  {Donner}}]{Mitra2017a}%
  \BibitemOpen
  \bibfield  {author} {\bibinfo {author} {\bibfnamefont {C.}~\bibnamefont
  {Mitra}}, \bibinfo {author} {\bibfnamefont {T.}~\bibnamefont {Kittel}},
  \bibinfo {author} {\bibfnamefont {A.}~\bibnamefont {Choudhary}}, \bibinfo
  {author} {\bibfnamefont {J.}~\bibnamefont {Kurths}}, \ and\ \bibinfo {author}
  {\bibfnamefont {R.~V.}\ \bibnamefont {Donner}},\ }\bibfield  {title}
  {\enquote {\bibinfo {title} {Recovery time after localized perturbations in
  complex dynamical networks},}\ }\href@noop {} {\bibfield  {journal} {\bibinfo
   {journal} {arXiv preprint arXiv:1704.06079}\ } (\bibinfo {year}
  {2017}{\natexlab{b}})}\BibitemShut {NoStop}%
\bibitem [{\citenamefont {Agresti}\ and\ \citenamefont
  {Coull}(1998)}]{Agresti1998}%
  \BibitemOpen
  \bibfield  {author} {\bibinfo {author} {\bibfnamefont {A.}~\bibnamefont
  {Agresti}}\ and\ \bibinfo {author} {\bibfnamefont {B.~A.}\ \bibnamefont
  {Coull}},\ }\bibfield  {title} {\enquote {\bibinfo {title} {{Approximate is
  Better than "Exact" for Interval Estimation of Binomial Proportion}},}\
  }\href@noop {} {\bibfield  {journal} {\bibinfo  {journal} {The American
  Statistician}\ }\textbf {\bibinfo {volume} {52}},\ \bibinfo {pages}
  {119--126} (\bibinfo {year} {1998})}\BibitemShut {NoStop}%
\bibitem [{Note1()}]{Note1}%
  \BibitemOpen
  \bibinfo {note} {Note that we applied a phase shift of $\protect \qopname
  \relax o{arcsin}\mathopen {}\left ( p/k \right )\mathclose {}$ to set the fix
  point to the origin.}\BibitemShut {Stop}%
\bibitem [{\citenamefont {Serdukova}\ \emph {et~al.}(2016)\citenamefont
  {Serdukova}, \citenamefont {Zheng}, \citenamefont {Duan},\ and\ \citenamefont
  {Kurths}}]{Serdukova2016Stochastic}%
  \BibitemOpen
  \bibfield  {author} {\bibinfo {author} {\bibfnamefont {L.}~\bibnamefont
  {Serdukova}}, \bibinfo {author} {\bibfnamefont {Y.}~\bibnamefont {Zheng}},
  \bibinfo {author} {\bibfnamefont {J.}~\bibnamefont {Duan}}, \ and\ \bibinfo
  {author} {\bibfnamefont {J.}~\bibnamefont {Kurths}},\ }\bibfield  {title}
  {\enquote {\bibinfo {title} {Stochastic basins of attraction for metastable
  states},}\ }\bibfield  {booktitle} {\emph {\bibinfo {booktitle} {Chaos: An
  Interdisciplinary Journal of Nonlinear Science}},\ }\href@noop {} {\bibfield
  {journal} {\bibinfo  {journal} {Chaos: An Interdisciplinary Journal of
  Nonlinear Science}\ }\textbf {\bibinfo {volume} {26}},\ \bibinfo {pages}
  {073117} (\bibinfo {year} {2016})}\BibitemShut {NoStop}%
\bibitem [{\citenamefont {Serdukova}\ \emph {et~al.}(2017)\citenamefont
  {Serdukova}, \citenamefont {Zheng}, \citenamefont {Duan},\ and\ \citenamefont
  {Kurths}}]{Serdukova2017Metastability}%
  \BibitemOpen
  \bibfield  {author} {\bibinfo {author} {\bibfnamefont {L.}~\bibnamefont
  {Serdukova}}, \bibinfo {author} {\bibfnamefont {Y.}~\bibnamefont {Zheng}},
  \bibinfo {author} {\bibfnamefont {J.}~\bibnamefont {Duan}}, \ and\ \bibinfo
  {author} {\bibfnamefont {J.}~\bibnamefont {Kurths}},\ }\bibfield  {title}
  {\enquote {\bibinfo {title} {Metastability for discontinuous dynamical
  systems under lévy noise: Case study on amazonian vegetation},}\ }\bibfield
  {booktitle} {\emph {\bibinfo {booktitle} {Scientific Reports}},\ }\href@noop
  {} {\bibfield  {journal} {\bibinfo  {journal} {Scientific Reports}\ }\textbf
  {\bibinfo {volume} {7}},\ \bibinfo {pages} {9336} (\bibinfo {year}
  {2017})}\BibitemShut {NoStop}%
\bibitem [{\citenamefont {Imkeller}\ and\ \citenamefont
  {Pavlyukevich}(2006)}]{Imkeller2006}%
  \BibitemOpen
  \bibfield  {author} {\bibinfo {author} {\bibfnamefont {P.}~\bibnamefont
  {Imkeller}}\ and\ \bibinfo {author} {\bibfnamefont {I.}~\bibnamefont
  {Pavlyukevich}},\ }\bibfield  {title} {\enquote {\bibinfo {title}
  {{L{\'{e}}vy flights: transitions and meta-stability}},}\ }\href@noop {}
  {\bibfield  {journal} {\bibinfo  {journal} {J. Phys. A. Math. Gen.}\ }\textbf
  {\bibinfo {volume} {39}},\ \bibinfo {pages} {L237--L246} (\bibinfo {year}
  {2006})}\BibitemShut {NoStop}%
\bibitem [{\citenamefont
  {Pavlyukevich}(2007{\natexlab{a}})}]{Pavlyukevich2007}%
  \BibitemOpen
  \bibfield  {author} {\bibinfo {author} {\bibfnamefont {I.}~\bibnamefont
  {Pavlyukevich}},\ }\bibfield  {title} {\enquote {\bibinfo {title}
  {{L{\'{e}}vy flights, non-local search and simulated annealing}},}\
  }\href@noop {} {\bibfield  {journal} {\bibinfo  {journal} {J. Comput. Phys.}\
  }\textbf {\bibinfo {volume} {226}},\ \bibinfo {pages} {1830--1844} (\bibinfo
  {year} {2007}{\natexlab{a}})}\BibitemShut {NoStop}%
\bibitem [{\citenamefont
  {Pavlyukevich}(2007{\natexlab{b}})}]{Pavlyukevich2007a}%
  \BibitemOpen
  \bibfield  {author} {\bibinfo {author} {\bibfnamefont {I.}~\bibnamefont
  {Pavlyukevich}},\ }\bibfield  {title} {\enquote {\bibinfo {title} {{Cooling
  down L{\'{e}}vy flights}},}\ }\href@noop {} {\bibfield  {journal} {\bibinfo
  {journal} {J. Phys. A Math. Theor.}\ }\textbf {\bibinfo {volume} {40}},\
  \bibinfo {pages} {12299--12313} (\bibinfo {year}
  {2007}{\natexlab{b}})}\BibitemShut {NoStop}%
\end{thebibliography}%

\end{document}